\documentclass[fleqn,twoside]{article}
\usepackage[headings]{espcrc2}

\readRCS
$Id: espcrc2.tex,v 1.2 2004/02/24 11:22:11 spepping Exp $
\usepackage{graphicx}
\usepackage[figuresright]{rotating}

\newcommand{\AmS}{{\protect\the\textfont2
  A\kern-.1667em\lower.5ex\hbox{M}\kern-.125emS}}

\title{\textbf{An Improved Efficient Key Distribution Mechanism for Large-Scale Heterogeneous Mobile Sensor Networks}}

\author{Ashok Kumar Das \address[MCSD]{International Institute of Information Technology, Bhubaneswar \\
Bhubaneswar 751013, India \\
E-mail: iitkgp\_akdas2006@yahoo.co.in}}

\runtitle{An Improved Efficient Key Distribution Mechanism}
\runauthor{A. K. Das}

\begin{document}

\begin{abstract}
Due to resource constraints of the sensor nodes, traditional public key cryptographic techniques are not feasible in most sensor network architectures. Several symmetric key distribution mechanisms are proposed for establishing pairwise keys between sensor nodes in sensor networks, but most of them are not scalable and also are not much suited for mobile sensor networks because they incur much communication as well as computational overheads. Moreover, these schemes are either vulnerable to a small number of compromised sensor nodes or involve expensive protocols for establishing keys. In this paper, we introduce a new scheme for establishing keys between sensor nodes with the help of additional high-end sensor nodes, called the \textit{auxiliary nodes}. Our scheme provides unconditional security against sensor node captures and high network connectivity. In addition, our scheme requires minimal storage requirement for storing keys mainly due to only a single key before deployment in each node in the sensor network, supports efficiently addition of new nodes after initial deployment and also works for any deployment topology.\\

\vspace{1pc}
\end{abstract}

\maketitle

\section{Introduction}
In a sensor network, many small computing nodes are deployed for the purpose of sensing data and then transmitting the data to nearby base stations. Public-key cryptosystems are not suited in such networks because of resource limitations of the nodes and also due to the vulnerability of physical captures of the nodes by the enemy. The symmetric ciphers such as DES~\cite{bk:02}, AES~\cite{cf:10}, RC5~\cite{cf:25} are the viable options for encrypting secret data. However, setting up symmetric keys among communicating nodes remains a challenging problem. A survey on sensor networks can be found in \cite{jr:06}.\\
\indent The topology of the network is in general dynamic, because the connectivity among the sensor nodes may vary with time due to several factors such as node addition, node departure, and the possibility of having mobile nodes. In our case, we assume that the nodes are highly mobile.\\
\indent Pairwise key establishment between neighboring sensor nodes in a sensor network is done by a protocol known as the \emph{bootstrapping protocol}. A bootstrapping protocol usually involves several steps. In the \emph{key pre-distribution phase}, each sensor node is loaded by a set of pre-distributed keys in its memory. This is done before deployment of the sensor nodes in a target field. After deployment, a \emph{direct key establishment (shared key discovery) phase} is performed by the nodes in order to establish direct pairwise keys between them. Two nodes $u$ and $v$ are called \emph{physical neighbors} if they are within the communication ranges of one another. They are called \emph{key neighbors} if they share one or more key(s) in their key rings $K_{u}$ and $K_{v}$. The nodes $u$ and $v$ can secretly and directly communicate with one another if and only if they are both physical and key neighbors. In this case we call $u$ and $v$ \emph{direct neighbors}. The \textit{path key establishment phase} is an optional stage and, if executed, adds to the connectivity of the network. In this phase, a secure path is discovered between two physical neighbors and a fresh pairwise key is sent securely along that path. \\
\indent The simplest solution is the use of a single mission key for the entire network. In this case, each node is given the same mission key before deployment in the network. After deployment any two neighbor nodes can communicate among each other using this key. Unfortunately, the compromise of even a single node in a network would reveal the secret key and thus allow decryption of all network traffic. SPINS~\cite{cf:06} is a protocol to provide pairwise key establishment using the base station as a KDC (key distribution center). However, this protocol incurs huge communication overhead and it is also vulnerable to single point failure. Several techniques~\cite{cf:01},~\cite{cf:02},~\cite{cf:04},~\cite{cf:07},~\cite{cf:08} are proposed in the literature in order to solve the bootstrapping problem. These schemes are not much suitable for mobile sensor networks because the bootstrapping procedure involves considerable communication and computational overheads and can not be repeated quite often. \\
\indent  The key pre-distribution scheme~\cite{cf:26} proposed for mobile heterogeneous wireless sensor networks is an improvement of the Dong and Liu's scheme~\cite{cf:18}. This scheme supports large-scale sensor networks and dynamic node addition efficiently after initial deployment. In this scheme, each regular sensor node requires to store equal number of keying information as in other schemes~\cite{cf:01,cf:02,cf:04,cf:18}.\\
\indent In this paper, we propose a new key pre-distribution mechanism suitable for highly mobile sensor networks. Our scheme requires minimal storage requirement due to storing a single key in each sensor's memory. Security analysis and performance evaluation illustrate that our scheme provides better performance than existing protocols, in terms of network connectivity, key storage overhead and network resilience against node capture attacks. Simulation results also demonstrate that our scheme is highly applicable for mobile wireless sensor networks.\\
\indent The rest of this paper is organized as follows. In Section 2, we briefly describe the Dong and Liu's scheme~\cite{cf:18}, which can be applied for highly mobile sensor nodes. Section 3 introduces our proposed scheme. In Section 4, we give a detailed analysis of our scheme with respect to network connectivity, resilience against node captures and overheads required by a sensor node in order to establish pairwise keys with its neighbor nodes. We discuss the simulation results of our scheme in Section 5. Section 6 compares the performances of our scheme with the previous schemes. Finally, we conclude the paper in Section 7.

\section{Overview of the Dong and Liu's Scheme}
 This scheme proposed by Dong and Liu~\cite{cf:18} recently, does not depend on the sensors' location information and can be used for the sensor networks with highly mobile sensor nodes. The main idea behind their approach is to deploy additional sensor nodes, called \textit{assisting nodes}, to help the pairwise key establishment between sensor nodes in a sensor network. Thus, they will not be useful for secure node-to-node communication in the network. The sensor nodes which are not assisting nodes, called as the \textit{regular sensor nodes}.\\
\indent In the \textit{key pre-distribution phase}, before deployment, the base station generates a unique master key $MK_{u}$ for every sensor node $u$ which will be shared between $u$ and the base station only. Every assisting node $i$ will get preloaded with a hash value $H(MK_{u} || i)$ for every regular sensor node $u$, where $H$ is a one-way hash function (for example, SHA-1~\cite{ms:01}) and "$||$" denotes the concatenation operation. If $n$ is the number of regular sensor nodes in the network, it is thus assumed that each assisting node will store all $n$ hash images.\\
\indent In the \textit{pairwise key establishment phase}, after deployment, every regular sensor node first discovers the assisting nodes as well as other regular sensor nodes in its communication range. If a regular sensor node $u$ needs to establish a pairwise key with its neighbor regular sensor node $v$, it will send a request to its every neighbor assisting node $i$. The request message includes the IDs of $u$ and $v$ and will be protected by the key $H(MK_{u} || i)$, which is already preloaded to the assisting node $i$. We note that, in this case, the assisting node acts as a KDC (key distribution center). After receiving such a request from $u$, the assisting node $i$ will generate a reply to $u$. The assisting node $i$ generates a random key $R$, makes two copies of it, one protected by $H(MK_{u} || i)$ (for node $u$) and other protected by $H(MK_{v} || i)$ (for node $v$) and finally sends these two protected copies to node $u$. Node $u$ then gets $R$ by decrypting the protected $R$ using the key $H(MK_{u} || i)$ and forwards another protected copy of $R$ to node $v$. Similarly, node $v$ retrieves $R$ by decrypting the protected $R$ using the key $H(MK_{v} || i)$. Thus, $u$ will receive a random key from every neighbor assisting neighbor node. If $\{ R_1$, $R_2$,$\ldots$, $R_l \}$ be such a set of all these random keys, the final key between $u$ and $v$ is computed as $k_{u,v} =$ $R_1 \oplus$ $R_2\oplus$ $\cdots$ $\oplus R_l$. We note that as long as at least one random key is secure, this final key $k_{u,v}$ will be safe.\\
\indent It may be possible that a regular sensor node could not find any assisting sensor node in its neighborhood. In that case, the \textit{supplemental key establishment phase} will be performed, where a regular sensor node may discover a set of assisting nodes with a certain number of hops in order establish a pairwise key with its neighbor node.\\
\indent This scheme guarantees high probability of establishing keys between regular sensor nodes and provides better security compared to those for the EG scheme~\cite{cf:01}, the $q$-composite scheme~\cite{cf:02}, the random pairwise keys scheme~\cite{cf:02} and the polynomial-pool based scheme~\cite{cf:04}. The main drawback of this scheme is that it does not support dynamic regular sensor node addition after initial deployment, which is considered as a very crucial issue in sensor networks.

\section{Our Proposed Scheme}
In this section, we describe the motivation behind development of our novel scheme. We then describe assumptions, notations and various phases of our scheme.

\subsection{Motivation}
Our scheme is motivated by the following considerations. In the Dong and Liu's scheme~\cite{cf:18}, it is necessary to store a large number of hash images of keys of the regular sensor nodes in each assisting node's memory. In particular, for a large-scale sensor network, an assisting node has thus to store huge number of hash values of keys of the regular sensor nodes.  Moreover, the assisting nodes are only be used for pairwise key establishment purpose between regular sensor nodes and hence they do not helpful for secure node-to-node communication as well as authentication in a sensor network. Thus, in case of node compromising they will not be helpful for secure communications. \\
\indent Though their scheme provides high network connectivity and much better resilience against node captures than the EG scheme~\cite{cf:01}, the $q$-composite scheme~\cite{cf:02}, the random pairwise keys scheme~\cite{cf:02} and the polynomial-pool based scheme~\cite{cf:04}, it can not still guarantee perfect security (i.e., unconditional security) against node captures even after enhancing their scheme by updating the keys for the regular sensor nodes in assisting nodes after key establishment. Thus, it introduces considerable computational overheads for the assisting sensor nodes in order to update those keys for better security. Another drawback of their scheme is that it does not support dynamic regular sensor node addition after initial deployment. Addition of new regular sensor nodes is extremely important for a sensor network, because sensor nodes may be compromised by an attacker or may be expired due to energy consumption, and as a result, we always expect to deploy some fresh nodes to join in the network in order to continue security services.  \\
\indent To overcome the above aforementioned problems, we propose a new key pre-distribution  scheme suited for highly mobile sensor nodes. As in~\cite{cf:26}, we make an assumption that a mobile node is aware of its physical movement. After each movement, a node will initiate a pairwise key establishment procedure in its new physical neighborhood. \\

\indent Our scheme has the following interesting properties:
\begin{itemize}
\item It provides high network connectivity.

\item It is unconditionally secure against node captures. This means that no matter how many nodes are compromised, an attacker will never compromise secret communications between non-compromised nodes in the network.

\item It requires minimal storage requirement for storing a single key in each node's memory.

\item It supports dynamic node addition efficiently after initial deployment.  
\end{itemize}

\subsection{Assumptions}
We list the following basic assumptions as used in~\cite{jr:14} for developing our protocol:
\begin{itemize}
\item \textbf{A1:} We consider a wireless sensor network consisting of two types of sensors: a small number of High-end sensors (H-sensors) and a huge number of Low-end sensors (L-sensors). The H-sensors are equipped with high power batteries, large memory storages and data processing capacities. They can execute relatively complicated numerical operations. On the other hand, the L-sensors are inexpensive and extremely resource constrainted (as in \cite{cf:17}). Each L-sensor has limited battery power, memory size, data processing capability and short radio transmission range. For example, the H-sensors can be PDAs and the L-sensors are the MICA2-DOT motes~\cite{ms:03}.

\item \textbf{A2:} H-sensors are equipped with tamper resistant hardware. In practical applications, it is true since only a small number of such nodes to be deployed in the network. Thus, all the keying information stored in each H-sensor will not be disclosed to an attacker even if he captures any H-sensor node in the network.

\item \textbf{A3:} Due to cost constraints, L-sensors are not equipped with tamper resistant hardware. We assume that an attacker can eavesdrop on all traffic, inject packets and reply old messages previously delivered. We further assume that if an attacker captures an L-sensor node, all the keying information it holds will also be compromised.

\item \textbf{A4: } After deployment, we assume that H-sensors will be static, whereas L-sensors are highly mobile nodes in the network.

\end{itemize}

\subsection{Notations}
Following is the convention we use to describe the protocol in this paper.
\begin{itemize}
\item $u\rightarrow v:$ $M$ refers to a message $M$ sent from a node $u$ to another node $v$.

\item $id_{u}: $ the unique identifier of a sensor node $u$, that is, a name of node $u$.

\item $MK_{u}: $ the master key of a sensor node $u$ to be shared with the base station only.

\item $SK: $ the special key shared by all the H-sensors and the base station.

\item $k_{u,v}$: a shared secret key between nodes $u$ and $v$.

\item $E_{k}(M): $ a message $M$ encrypted using key $k$.

\item $MAC_{k}(M): $ a message authentication code (MAC) for the message $M$, under the key $k$.

\item $PRF_{k}(M): $ output of a pseudo-random function (PRF) for the message $M$, under the key $k$.

\item $RN_{u}: $ a random nonce generated by the sensor node $u$. Nonce is a one-time random bit-string, usually used to achieve freshness.

\item $A || B: $ data $A$ concatenates with data $B$.
\end{itemize}

\subsection{Protocol Description}
The main idea behind our scheme is to deploy a small number of additional H-sensor nodes along with a large number of L-sensor nodes in the network, so that H-sensors can help in pairwise key establishment procedure between sensor nodes. The L-sensor nodes are to be deployed randomly in a target field, whereas the H-sensor nodes will be deployed to their expected locations. After deployment, sensor nodes will form multi-hop infrastructure-less wireless communication between each other and finally communicate with the base station. The base station is most powerful node and it can be located either at the center or at a corner of the target field. Let $m$ be the number of H-sensor nodes and $n$ the number of L-sensor nodes where $n >> m$. For convenience, we call the H-sensors as the \textit{auxiliary nodes} because they will help for pairwise key establishment procedure and the L-sensors as the \textit{regular sensor nodes}. Our protocol has the following phases.

\subsubsection{Key Pre-Distribution Phase}
This phase is done in offline by a key setup server before deployment of the nodes in a target field. It consists of the following steps:
\begin{itemize}
\item \textbf{\textit{Step-1:}} For each auxiliary sensor node $AS_{i}$, the setup server assigns a unique identifier, say $id_{AS_i}$. Similarly, for each regular sensor node $u$, the setup server also assigns a unique identifier, say $id_{u}$.

\item \textbf{\textit{Step-2:}} The setup server randomly generates a special key $SK$ which will be given to all the auxiliary sensor nodes and the base station.  

\item \textbf{\textit{Step-3:}} For each deployed regular sensor node $u$, the setup server randomly generates a master key $MK_u$ which will be shared with the base station only. This master key is computed using the special key $SK$ and the identifier $id_u$ as $MK_u =$ $PRF_{SK}(id_{u})$, where $PRF$ is a pseudo random function proposed by Goldreich et al.~\cite{jr:08}.

\item \textbf{\textit{Step-4:}} Finally, the setup server loads $(i)$ the identifier $id_{AS_i}$ and $(ii)$ the special key $SK$ to the memory of each auxiliary sensor node $AS_i$. Each regular sensor node $u$ is loaded with the following information: $(i)$ its own identifier $id_{u}$ and $(ii)$ its own master key $MK_{u}$ before deployment.  
\end{itemize}

\subsubsection{Direct Key Establishment Phase}
In this phase, we consider the following two cases.
\begin{description}
\item {\textbf{Case 1: Key establishment between regular sensor nodes} }\\

After deployment of the nodes, every regular sensor node discovers an auxiliary sensor node in its communication range. If a regular sensor node $u$ wants to establish a secret pairwise key with its neighbor regular sensor node $v$, node $v$ will send a request to its auxiliary node. Let $AS_i$ be an auxiliary node discovered by the node $v$ within its neighborhood. Then the auxiliary node $AS_i$ will generate a random key $k_{u,v}$, create two protected copies of it, one for $u$ and other for $v$, and send a reply containing these two protected copies of $k_{u,v}$ to node $v$. Upon receiving such a reply from $AS_i$, node $v$ will retrieve $k_{u,v}$ and forward other copy to $u$. $u$ will also retrieve $k_{u,v}$. Finally, $u$ and $v$ will store $k_{u,v}$ for their future secret communications.\\
\indent ~~~~This phase for establishing pairwise keys between $u$ and $v$ is summarized below:
\begin{enumerate}
\item Node $u$ generates a random nonce $RN_{u}$ and sends the following message to node $v$: \\
      $u\rightarrow v:$ $(id_{u} || RN_{u})$. 

\item Node $v$ also generates a random nonce $RN_{v}$ and sends the following message to its auxiliary node $AS_{i}$:\\
     $v\rightarrow AS_i:$ $T=$$(id_{u} || id_{v} || RN_{u} || RN_{v})$ $||$ $MAC_{MK_{v}}(T)$. 

\item The auxiliary node $AS_i$ first computes the master key of $v$ as $MK_{v}$ $= PRF_{SK}(id_{v})$ and then computes $MAC_{MK_{v}}(T)$. If the computed MAC and the received MAC are equal, $v$ is considered as a legitimate node. If the authentication is successful, then only the auxiliary node $AS_i$ computes the master key of $u$ as $MK_{u} = $ $PRF_{SK}(id_u)$. Finally, the auxiliary node $AS_i$ generates a random key $k_{u,v}$ and sends two copies of it, one protected by $MK_u$ (for node $u$) and another protected by $MK_v$ (for node $v$) to node $v$ as follows:\\
   $AS_i\rightarrow v:$ $(E_{MK_u}(k_{u,v} \oplus id_u \oplus RN_u), E_{MK_v}(k_{u,v} \oplus id_v \oplus RN_v))$. 

\item Node $v$ decrypts $E_{MK_v}(k_{u,v} \oplus id_v \oplus RN_v)$ using its own master key $MK_{v}$, retrieves $k_{u,v}$ using its own identifier and its own random nonce as $k_{u,v} =$ $(k_{u,v} \oplus id_v \oplus RN_v$) $\oplus (id_v \oplus RN_v)$ and forwards other protected copy of $k_{u,v}$ to node $u$:\\
   $v\rightarrow u:$ $E_{MK_u}(k_{u,v} \oplus id_u \oplus RN_u)$.
\end{enumerate}

\noindent Similarly, after receiving the last message from $v$, $u$ retrieves $k_{u,v}$ using its own master key $MK_{u}$. \textit{The auxiliary sensor node $AS_i$ discards the computed master keys $MK_u$ (for regular sensor node $u$) as well as $MK_v$ (for regular sensor node $v$) and the generated key $k_{u,v}$ from its memory}. The special key $SK$ is secure, since the auxiliary nodes are equipped with tamper resistant hardware. Therefore, based on the security of the PRF function~\cite{jr:08}, even an attacker knows a master key of a captured regular sensor node, it is computationally infeasible to compute the special key $SK$. We note that each auxiliary node acts as a KDC. We also observe that the transaction between two nodes is uniquely determined by attaching the random nonce of a node in the message by another node. 

\item{\textbf{Case 2: Key establishment between auxiliary nodes and regular sensor nodes}}\\

 In case, if a regular sensor node, say $u$ wants to establish a pairwise key with its neighbor auxiliary node, say $AS_j$, $u$ simply sends a request message containing its own identifier $id_u$ to $AS_j$. After receiving such request from the node $u$, $AS_j$ generates a random key $k_{u,AS_j}$. After that $AS_j$ computes the master key of $u$ using the special key $SK$ and the identifier $id_u$ of node $u$ as $MK_u =$ $PRF_{SK}(id_u)$, makes a protected copy of the generated random key as $E_{MK_u}(k_{u,AS_j})$ and sends this to node $u$. After receiving such a message, $u$ decrypts this protected key using its own master key $MK_u$. In this way, $u$ and $AS_j$ will establish a pairwise key and store this key for secret communications in future. \textit{The auxiliary sensor node $AS_j$ also discards the computed master key $MK_u$ for the node $u$ from its memory}.
\end{description}

\subsubsection{Supplemental Key Establishment Phase}
Though our later analysis in Section 4.1 shows that even for a small fraction of the auxiliary sensor nodes it is possible to achieve a high probability of establishing pairwise keys using our direct key establishment phase, it is still possible that a regular sensor node cannot find an auxiliary sensor node in its neighborhood because the accurate deployment of the auxiliary nodes in their expected locations may not be guaranteed in some scenarios. In this situation, we have the \textit{supplemental key establishment phase}. In this phase, a regular sensor node discovers an auxiliary sensor node that is no more than $h$ hops away from itself. This can be also easily achieved by having $u$'s neighbors to help collecting the id of an auxiliary node in their neighborhood. Once an auxiliary sensor node is discovered, the remaining steps are similar to the direct key establishment phase. We note that the discovery and usage of an auxiliary node using multiple hops introduces additional communication overhead. However, we will show in Section 4.1 that by using only one hop, it is sufficient to achieve a very high probability of establishing pairwise keys between nodes. Also, the need for this phase will not likely be invoked frequently and hence we believe that this will not be a big problem. 

\subsubsection{Dynamic Node Addition Phase}
Sometimes nodes may be faulty due to battery-energy consumption, malfunctioning, etc. or compromised by an attacker by capturing the nodes. Therefore, it is necessary to redeploy some new sensor nodes to replace those faulty or compromised sensor nodes, or to deploy fresh nodes in the sensor network.\\
\indent In order to deploy a new regular sensor node $u$, the key setup server has to generate a unique identifier $id_u$ and a master key $MK_u$ which is computed using the special key $SK$ and the identifier $id_u$ as $MK_u =$ $PRF_{SK}(id_u)$ and loads these informations to its memory.\\
\indent In order to deploy a new auxiliary sensor node, say $AS_i$, the key setup server assigns a unique identifier, say $id_{AS_i}$. After that the setup server loads this identifier $id_{AS_i}$ and the previous special key $SK$ which is already with all the deployed auxiliary sensor nodes in the network.\\
\indent After deployment of such new nodes in the network, they establish pairwise keys using our direct key establishment phase and if required, using the supplemental key establishment phase. Thus, we see that dynamic node addition in our protocol is done efficiently.

\section{Analysis of Our Scheme}
In this section, we analyze the network connectivity of our scheme i.e., the probability of establishing pairwise keys between neighbor nodes in the network. We then analyze the resilience against node capture. Finally, we discuss overheads required in our scheme in order to establish pairwise keys between neighbor nodes.

\subsection{Network Connectivity}
For simplicity, we assume that nodes are evenly distributed in the target field. Let $m$ be the number of auxiliary sensor nodes and $n$ the number of regular sensor nodes in the target field. Let $d$ be the average number of neighbor nodes of each node. Let $u$ and $v$ two neighbor nodes. We assume that $p$ denotes the overall probability that any two neighbor nodes $u$ and $v$ can establish a pairwise key between them in the direct key establishment phase and $p_1$ the probability of establishing pairwise keys between two neighbor sensor nodes $u$ and $v$ using one hop auxiliary sensor node from $v$ in the supplemental key establishment phase. \\

\indent In order to establish a pairwise key between two neighbor regular sensor nodes $u$ and $v$, if $u$ initiates a request to establish a pairwise key with $v$, $v$ is required to communicate with an auxiliary sensor node in its communication range. The probability that an auxiliary node $AS_i$ is not in the neighborhood of the regular sensor node $v$ can be estimated by $1 -$ $\frac{d}{m+n}$. Since there are $m$ auxiliary nodes in the network, the probability that the regular sensor node $v$ fails to discover any auxiliary node $AS_i$ in its neighborhood can be estimated by $\left( {1 - \frac{d}{m+n}} \right) ^{m}$. As a result, the probability of establishing a pairwise key between two neighbor regular sensor nodes $u$ and $v$ is given by
\begin{equation}
p' = 1 - \left( {1 - \frac{d}{m+n}} \right)^{m}.
\end{equation}

\indent Since there are $n$ regular sensor nodes and $m$ auxiliary sensor nodes in the network, the total direct communication links is $\frac{\sum_{i=1}^{m+n} d}{2} =$ $\frac{(m+n) d}{2}$, since there are $d$ average number of neighbor nodes for each node. We note from our direct key establishment phase that each auxiliary node can establish a secret key with its neighbor regular sensor nodes. Since there are $m$ auxiliary nodes in the network, secure communication links due to key establishment between auxiliary nodes and their neighbor regular sensor nodes becomes $\frac{m d}{2}$. Again, we have another $\frac{n d}{2} \times p'$ secure links in the network due to key establishment between regular sensor nodes with the help of auxiliary nodes. As a result, we obtain $(\frac{m d}{2} + \frac{n d}{2} \times p')$ secure links out of total $\frac{(m+n) d}{2}$ links. Hence, the required overall network connectivity is given by
\begin{equation}
p = \frac{\frac{m d}{2} + \frac{n d}{2} \times p'}{\frac{(m+n) d}{2}}
   = \frac{m + n \times p'}{m + n}.
\end{equation}

\indent Figure 1 shows the relationship between the probability $p$ of establishing keys between neighbor nodes versus the fraction of auxiliary sensor nodes $\frac{m}{n}$. From this figure, we observe that $p$ increases faster as $d$ increases. Moreover, for a small fraction of auxiliary sensor nodes $\frac{m}{n}$, our scheme also guarantees a high probability of establishing keys between neighbor sensor nodes. 

\begin{figure}[htbp]
\centering
\includegraphics[scale=0.29, angle=270]{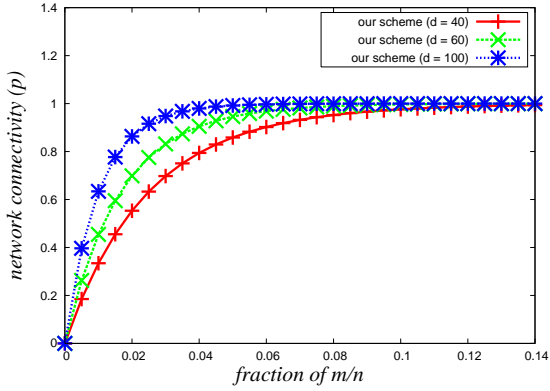}
\caption{Direct network connectivity $p$ of our scheme vs. the fraction of auxiliary sensor nodes $\frac{m}{n}$. } \label{Figure:fig1}
\end{figure}

\begin{figure}[htbp]
\centering
\includegraphics[scale=0.29, angle=270]{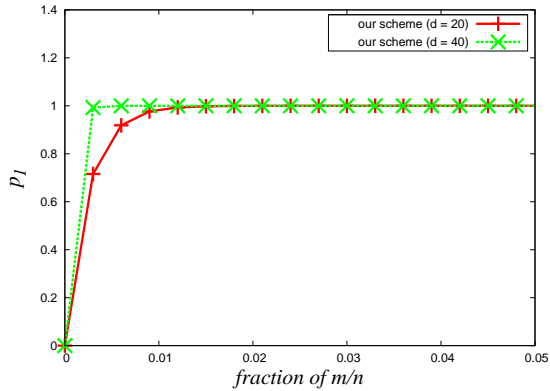}
\caption{Network connectivity $p_1$ vs. the fraction of auxiliary sensor nodes $\frac{m}{n}$ during one-hop supplemental key establishment phase of our scheme.} \label{Figure:fig2}
\end{figure}
\par
\indent From the supplemental key establishment phase, we notice that if a sensor node $v$ can not discover any auxiliary sensor node in its neighborhood, it will discover an auxiliary sensor node within $h$-hop range from it. Once an auxiliary sensor node is discovered, $v$ can establish a pairwise key with its neighbor node $u$. We now consider the case $h =$ $1$. The probability that none of the neighbors of $v$ can discover an auxiliary sensor node can be estimated by $\left( \left( 1 - \frac{d}{m+n} \right)^{m} \right)^{d}$. As a result, we have,
\begin{equation}
p_{1} = p + (1 - p) \times \! \left( 1 - \left( 1 - \frac{d}{m+n} \right)^{md} \right).
\end{equation}
\indent Figure 2 illustrates the relationship between the probability $p_1$ of establishing pairwise keys between neighbor nodes versus the fraction of auxiliary sensor nodes $\frac{m}{n}$. We note that once a regular sensor node discovers an auxiliary sensor node within one-hop neighbors of it, the probability of establishing pairwise keys between neighbor nodes becomes very high. As a result, we observe that one needs not to apply more than one-hop supplemental key establishment phase after the direct key establishment phase.

\subsection{Resilience against Node Capture}
The resilience against node capture attack of a scheme is measured by estimating the fraction of total secure communications that are compromised by a capture of $c$ nodes \textit{not including} the communication in which the compromised nodes are directly involved. In other words, we want to find out the effect of $c$ sensor nodes being compromised on the rest of the network. For example, for any two non-compromised sensor nodes $u$ and $v$, we have to find out what is the probability that the adversary can decrypt the secret communications between $u$ and $v$ when $c$ sensor nodes are already compromised ? \\
\indent From our key pre-distribution phase described in subsection 3.4.1, we see that each auxiliary sensor node is given the special key $SK$. This special key $SK$ is secure, because all the auxiliary sensor nodes are equipped with tamper resistant hardware. As a result, based on the security of the PRF function~\cite{jr:08}, it is computationally infeasible to recover the special key $SK$ from the master key $MK_u =$ $PRF_{SK}(id_u)$ of a compromised regular sensor node $u$. We also notice from our direct key establishment phase as well as supplemental key establishment phase that each pairwise key between nodes are generated randomly. Thus, no matter how many nodes are captured, an attacker knows nothing about the secret communications between non-compromised nodes. In other words, no matter how many nodes are captured, non-compromised nodes can still communicate with $100\%$ secrecy. In this way, our scheme guarantees \textit{unconditional security against node captures attack}.  

\subsection{Overheads}
Our scheme only requires a single master key $MK_u$ to be stored in each regular sensor node $u$ and the special key $SK$ in each auxiliary sensor node $AS_i$. \\
\indent  Our scheme involves small computation overheads. A regular sensor node has to perform one MAC operation and one symmetric key decryption. \\
\indent We note from our direct key establishment phase that the communication is in one-hop range only. Though multi-hop communication will introduce some communication overheads, we see from our analysis that the supplemental key establishment can be done using only one-hop procedure. Moreover, such phase needs not to be involved frequently, since direct key establishment phase will ensure high network connectivity.

\section{Simulation Results}
In this section, we describe the simulation results of network connectivity of our scheme.\\
\indent We have implemented our proposed scheme for static networks as well as mobile networks. We have considered the following parameters for simulation of network connectivity of our scheme as follows:
\begin{itemize}
\item The number of regular sensor nodes in the network is $n =$ $5000$.

\item The number of auxiliary sensor nodes is $m \le 500$.

\item The average number of neighbor nodes for each node is $d =$ $80$.

\item The communication range of each sensor node is $\rho =$ $30$ meters.

\item The area $A$ of the target field is taken so that  the maximum network size becomes $n =$ $\frac{(d+1) \times A}{\pi \rho^{2}}$.
\end{itemize}

\noindent We assume that the target field is a flat region. The regular sensor nodes are randomly deployed in the target field. For deployment of auxiliary sensor nodes, we logically divide the target field into $c \times c$ cells, where $c =$ $\lceil \sqrt{m} \, \rceil$. For each cell, an auxiliary sensor node is to be deployed randomly in that cell. \\
\begin{figure}[htbp]
\centering
\includegraphics[scale=0.29, angle=270]{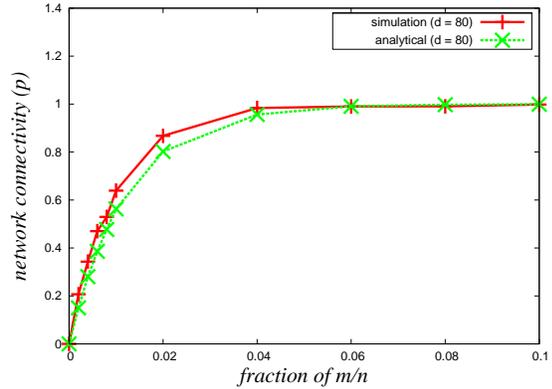}
\caption{Simulation vs. analytical results of direct network connectivity $p$ of our scheme, with $n =$ $5000$, $m \le$ $500$ and $d =$ $80$, under static network.} \label{Figure:fig3}
\end{figure}
\indent In our simulation, we only consider the direct key establishment phase in which if a regular sensor node $u$ wants to establish a pairwise key with its neighbor regular sensor node $v$, node $v$ will send a request to an auxiliary sensor node within its neighborhood. Once such an auxiliary sensor node is discovered by $v$, nodes $u$ and $v$ establish a pairwise key between them. \\
\indent The simulation results versus the analytical results of direct network connectivity $p$ of our scheme are plotted in Figure 3. In this figure, we assume that all the nodes are static. From this figure we observe that our simulation results closely tally with the analytical results.    

\begin{figure}[htbp]
\centering
\includegraphics[scale=0.29, angle=270]{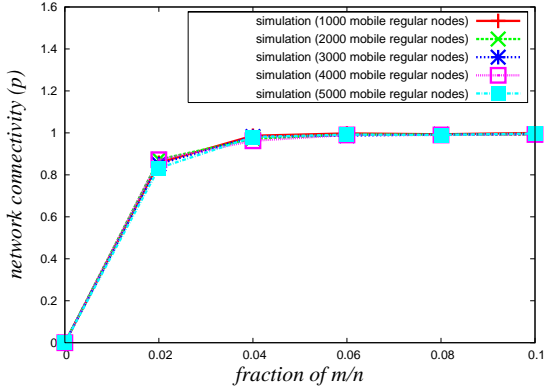}
\caption{Simulation results of direct network connectivity $p$ of our scheme for mobile regular sensor nodes under different situations, with $n =$ $5000$, $m \le$ $500$ and $d =$ $80$. } \label{Figure:fig4}
\end{figure}

\indent We have also simulated the direct network connectivity $p$ of our scheme under mobile environment. We assume that after deployment the auxiliary sensor nodes remain static, whereas the regular sensor nodes will be highly mobile. We have considered that any regular sensor node can move from its current location in the target field within the range of twice its communication range. The simulation results for mobile regular nodes are shown in Figure 4. From this figure, we see that our scheme is much suitable for highly mobile sensor networks, because no matter how many how many regular sensor nodes are mobile in the network, the overall network connectivity does not degrade.

\section{Comparison with Previous Schemes}
In this section, we compare the performances of our proposed method with those for the previous schemes, namely, the EG scheme~\cite{cf:01}, the $q$-composite scheme~\cite{cf:02}, the polynomial-pool based scheme~\cite{cf:04} and the Dong and Liu's scheme~\cite{cf:18}. The reason for considering those schemes is that there is no restriction in applying those schemes on mobile sensor networks.\\
\begin{figure}[htbp]
\centering
\includegraphics[scale=0.29,angle=270]{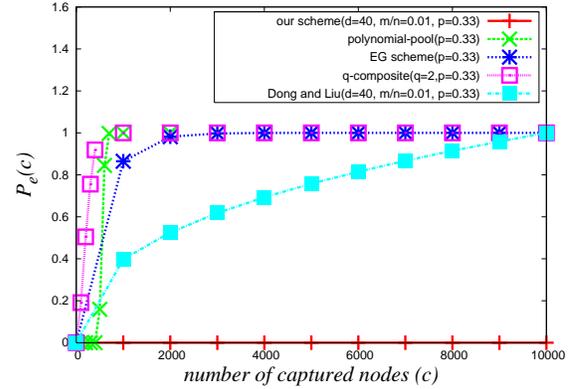}
\caption{Comparison of resilience against node captures between different schemes. $P_e(c)$ denotes the fraction of compromised links between non-compromised sensor nodes  after capturing $c$ sensor nodes in the network and $p$ denotes the probability of establishing pairwise keys between neighbor sensor nodes.} \label{Figure:fig5}
\end{figure}
Figure 5 shows the relationship between the fraction of compromised links between non-compromised nodes versus the number of nodes capture. For the EG scheme~\cite{cf:01}, the $q$-composite scheme~\cite{cf:02} and the polynomial-pool based scheme~\cite{cf:04}, we assume that each node has available storage of holding $200$ cryptographic keys in its memory. The network connectivity probability $p$ is taken as $p =$ $0.33$ with suitable values of the parameters for the different schemes. We clearly see that our scheme provides significantly much better security against node captures than the EG scheme, the $q$-composite scheme and the polynomial-pool based scheme. Moreover our scheme has better security than the Dong and Liu's scheme. \\
\indent 
In our proposed scheme, only a single master key is stored in each regular sensor node and also a single special key $SK$ is stored in each auxiliary node to achieve high performance. Thus, no matter how many sensor nodes are in the network, our scheme requires only to store a single key, which is extremely memory efficient for the large-scale wireless sensor networks. In the Dong and Liu's scheme, though each regular sensor node needs to store a single key, but each assisting node needs to store all the hash images of keys of regular sensor nodes. On the other hands, the EG scheme, the $q$-composite scheme and the polynomial-pool based scheme require considerable storage requirements for achieving high network performances. \\
\indent The existing schemes (the EG scheme, the $q$-composite scheme and the polynomial-pool based scheme) require considerable communication as well as computational overheads in order to establish pairwise keys between neighbor nodes. In these schemes, the path key establishment procedure is a complicated one and requires a lot of communication and computational overheads. In our scheme as well as Dong and Liu's scheme, only a few number of symmetric key operations and PRF operations (in our scheme) or hash operations (in Dong and Liu's scheme) are required in order to establish a pairwise key between neighbor nodes. Zhu et al.~\cite{jr:17} pointed out that due to the computational efficiency of PRF function, the computational overhead of PRF function is negligible. As a result, the actual computational overhead is low in our scheme compared to that for the Dong and Liu's scheme.

\section{Conclusion}
We have proposed a new key pre-distribution scheme suitable for highly mobile sensor nodes with the help of additional auxiliary sensor nodes. Our scheme provides very high network connectivity than the previous schemes. Our scheme always guarantees unconditional security against node capture attack. It supports dynamic node addition efficiently compared to the previous schemes. Overall, our scheme has better trade-off among network connectivity, security, storage, communication and computation overheads compared to those for the previous schemes. In addition, our scheme also works for any deployment topology.

\bibliographystyle{plain}
\bibliography{sensor}

\noindent \textbf{Author Biography} \\

\textbf{Ashok Kumar Das} is currently working as an Assistant Professor in the International Institute of Information Technology, Bhubaneswar 751013, India.  He has submitted Ph.D. Thesis in the Department of Computer Science and Engineering of the Indian Institute of Technology, Kharagpur 721 302, India. He received the M.Tech. degree in Computer Science and Data Processing from the Indian Institute of Technology, Kharagpur, India, in 2000. He also received the M.Sc. degree in Mathematics from the Indian Institute of Technology, Kharagpur, India, in 1998.  Prior to join in Ph.D., he worked with C-DoT (Centre for Development of Telematics), a premier telecom technology centre of Govt. of India at New Delhi, India from March 2000 to January 2004. During that period he worked there as a Research Engineer on various projects in the fields of SS7 (Signaling System No. 7) protocol stack, GSM (Global System for Mobile Communications) and GPRS (General Packet Radio Services). \\
\indent He received the INSTITUTE SILVER MEDAL for his first rank in M.Sc. from the Indian Institute of Technology, Kharagpur, India in 1998. He has seventh All India  Rank in the Graduate Aptitude Test in Engineering (GATE) Examination in 1998. He received the DIVISIONAL AWARD for his individual excellence in development of SS7 protocol stack from C-DoT, New Delhi, India in 2003. He received a `Certificate of Special Mention' for the best paper award in the First International Conference on Emerging Applications of Information Technology (EAIT 2006) in 2006 and also a best paper award in the International Workshop on Mobile Systems (WoMS) in 2008. His current research  interests include cryptography and  wireless sensor network security.
\end{document}